\documentclass[a4paper,12pt]{article}
\pdfoutput=1
\usepackage{graphicx,rotating,colortbl}
\usepackage{hyperref}\usepackage{slashed}
\usepackage{charter}

\hypersetup{colorlinks,bookmarksopen,bookmarksnumbered,
linkcolor=blus,pdfstartview=FitH,urlcolor=rossos,citecolor=verde}

\newcommand{\fig}[1]{~\ref{fig:#1}}

\usepackage{youngtab}
\usepackage{multicol}
\usepackage{color}
\definecolor{rosso}{cmyk}{0,1,1,0.4}
\definecolor{rossos}{cmyk}{0,1,1,0.55}
\definecolor{rossoc}{cmyk}{0,1,1,0.2}
\definecolor{blu}{cmyk}{1,1,0,0.3}
\definecolor{blus}{cmyk}{1,1,0,0.6}
\definecolor{bluc}{cmyk}{1,1,0,0.1}
\definecolor{verde}{cmyk}{0.92,0,0.59,0.25}
\definecolor{verdec}{cmyk}{0.92,0,0.59,0.15}
\definecolor{verdes}{cmyk}{0.92,0,0.59,0.4}

\newcommand{\Q}{{\cal Q}}
\newcommand{\G}{{\cal G}}

\newcommand{\eq}[1]{~{\rm (\ref{eq:#1})}}

\newcommand{\GeV}{\,{\rm GeV}}
\newcommand{\TeV}{\,{\rm TeV}}
\newcommand{\cm}{\,{\rm cm}}
\newcommand{\Tr}{\,{\rm Tr}}

\def\circa#1{\,\raise.3ex\hbox{$#1$\kern-.75em\lower1ex\hbox{$\sim$}}\,}

\newcommand{\beq}{\begin{equation}}
\newcommand{\eeq}{\end{equation}}
\newcommand{\mb}[1]{\mbox{\boldmath $#1$}}

\newcommand{\bea}{\begin{eqnarray}}
\newcommand{\eea}{\end{eqnarray}}
\newcommand{\be}{\begin{equation}}
\newcommand{\ee}{\end{equation}}
\font\tenrsfs=rsfs10 at 12pt
\font\sevenrsfs=rsfs7
\font\fiversfs=rsfs5
\newfam\rsfsfam
\textfont\rsfsfam=\tenrsfs
\scriptfont\rsfsfam=\sevenrsfs
\scriptscriptfont\rsfsfam=\fiversfs
\def\mathscr#1{{\fam\rsfsfam\relax#1}}

\def\Lag{\mathscr{L}}

\def\circa#1{\,\raise.3ex\hbox{$#1$\kern-.75em\lower1ex\hbox{$\sim$}}\,}
\makeatletter

\def\hhref#1{\href{http://arxiv.org/abs/#1}{arXiv:#1}} % in bibliography

\def\art{\@ifnextchar[{\eart}{\oart}}
\def\eart[#1]#2#3#4#5#6{{\rm #2}, {\rm #3 #4} {\rm (#6) #5} [\hhref{#1}]}
\def\hepart[#1]#2{{\rm #2, \hhref{#1}}}
\newcommand{\oart}[5]{{\rm #1}, {\em #2 \bf #3} {\rm (#5) #4}}

\newcounter{alphaequation}[equation]
\def\thealphaequation{\theequation\hbox to
0.6em{\hfil\alph{alphaequation}\hfil}}
\def\eqnsystem#1{
\def\@eqnnum{{\rm (\thealphaequation)}}
\def\@@eqncr{\let\@tempa\relax \ifcase\@eqcnt \def\@tempa{& & &} \or
  \def\@tempa{& &}\or \def\@tempa{&}\fi\@tempa
  \if@eqnsw\@eqnnum\refstepcounter{alphaequation}\fi
\global\@eqnswtrue\global\@eqcnt=0\cr}
\refstepcounter{equation} \let\@currentlabel\theequation \def\@tempb{#1}
\ifx\@tempb\empty\else\label{#1}\fi
\refstepcounter{alphaequation}
\let\@currentlabel\thealphaequation
\global\@eqnswtrue\global\@eqcnt=0 \tabskip\@centering\let\\=\@eqncr
$$\halign to \displaywidth\bgroup \@eqnsel\hskip\@centering
$\displaystyle\tabskip\z@{##}$&\global\@eqcnt\@ne
\hskip2\arraycolsep\hfil${##}$\hfil& \global\@eqcnt\tw@\hskip2\arraycolsep
$\displaystyle\tabskip\z@{##}$\hfil
\tabskip\@centering&\llap{##}\tabskip\z@\cr}
\def\endeqnsystem{\@@eqncr\egroup$$\global\@ignoretrue} \makeatother

\newcommand{\SU}{\,{\rm SU}}

\begin{document}

\vspace{2cm}

\begin{center}
{\LARGE \bf \color{rossos}
%The Minimal Scale-Invariant Standard Model}\\[1.5cm]
Dynamical generation of the weak and\\[3mm] Dark Matter scales from strong interactions}\\[1.5cm]

{\bf Oleg Antipin$^{a}$, Michele Redi$^{a}$, Alessandro Strumia$^{b}$}  
\\[7mm]

{\it $^a$INFN, Sezione di Firenze, Via G. Sansone, 1; I-50019 Sesto Fiorentino, Italy}\\[2mm]
{\it $^b$ Dipartimento di Fisica dell'Universit{\`a} di Pisa and INFN, Italy\\
and National Institute of Chemical Physics and Biophysics, Tallinn, Estonia}

\thispagestyle{empty}

\vspace{1.5cm}
{\large\bf\color{blus} Abstract}
\begin{quote}
Assuming that mass scales arise in nature only via dimensional transmutation,
we extend the dimension-less Standard Model by adding vector-like fermions charged under a new strong gauge interaction.
Their non-perturbative dynamics generates
a mass scale that is transmitted to the elementary Higgs boson
by electro-weak gauge interactions.  
In its minimal version the model has the same number of parameters as the Standard Model,
predicts that the electro-weak symmetry gets broken,
predicts new-physics in the multi-TeV region and is compatible with all existing bounds,
provides two Dark Matter candidates stable thanks to accidental symmetries:
a composite scalar in the adjoint of  $\SU(2)_L$ and a composite singlet fermion;
their thermal relic abundance is predicted to be
comparable to the measured cosmological DM abundance.
Some models of this type allow for extra Yukawa couplings;
DM candidates remain even if explicit masses are added.
\end{quote}

\bigskip
%\bigskip
%\bigskip

%\thispagestyle{empty}
\end{center}
\begin{quote}
{\large\noindent\color{blus} 
}

\end{quote}
\vspace{-1.5cm}

\tableofcontents

\newpage

\setcounter{footnote}{0}

\section{Introduction}

The idea that the weak scale could be dynamically generated from strong interactions has a long history.
Originally, techni-color models were developed as an alternative to the Higgs:
the weak interactions of the techni-quarks ${\cal Q}$ were chosen so that their condensates 
would break the SM electro-weak group and the weak scale was the techni-color scale.
This scenario was disfavoured by flavour and precision data even before the first LHC run,
where  the Higgs and no new physics was observed.

Later, strong dynamics was invoked to generate a composite or partially-composite Higgs, although realising complete 
models is so complicated that model-building is usually substituted by postulating effective Lagrangians with the needed properties.

Recently, models where new strong dynamics does not break the electro-weak symmetry nor provide a composite Higgs
have been considered in the literature, just because they are simple, 
phenomenologically viable and lead to interesting 
LHC phenomenology~\cite{Sundrum}. With abuse of language we use the old name `techni-color'.
In this paper we show that these models 
\begin{enumerate}
\item provide Dark Matter candidates;
\item provide a dynamical origin for the electro-weak scale, 
if we adopt the  scenario of `finite naturalness'~\cite{FN,Bardeen,agrav}.
\end{enumerate}

Point 2 amounts to assuming that quadratically divergent corrections to the Higgs mass have no physical meaning and can be ignored,
possibly because the fundamental theory does not contain any mass term~\cite{agrav}.
In this context, dynamical generation of the weak scale via dimensional transmutation has been 
realised with weakly-coupled dynamics, in models where an extra scalar $S$ has interactions that drive its quartic $\lambda_S |S|^4$ 
negative around  or above the weak scale: $S$ acquires a vev at this scale, and its interaction
$\lambda_{HS}|H|^2 |S|^2$ effectively becomes a Higgs mass term, $m^2=\lambda_{HS}\langle S \rangle^2$~\cite{prev}.
A related possibility is that the scalar $S$ is interacting with techni-quarks~\cite{Raidal} or charged under a techni-color gauge group~\cite{Lindner}  
and again $S$ acquires a vev or forms a condensate.  In all these models $\langle S\rangle$ can be pushed arbitrarily above the weak scale 
by making $\lambda_{HS}$ arbitrarily small, leaving no observable signals.

\medskip

We here consider simple models without any extra scalar $S$ beside the Higgs doublet $H$.
The SM is extended by adding a gauge group $G_{\rm TC}$ (for example $\SU(N)$) and  techni-quarks ${\cal Q}_L$
charged under the SM, as well as the corresponding ${\cal Q}_R$ in the conjugated representations of the gauge group
$G_{\rm SM}\otimes G_{\rm TC}$, so that  $\Q_L \oplus {\Q}_R$ is vectorial. 
As a consequence the condensate $\langle\Q_L\Q_R\rangle$ transforms as a singlet of  $G_{\rm SM}$ and does not break it.

The  techni-quarks have no mass terms because of our assumption that only dimension-less couplings exist\footnote{Relaxing this hypothesis allows other interesting possibilities for Dark Matter that will be discussed in a separate publication  \cite{compositeDM}.};
for certain assignments of their gauge quantum numbers,
techni-quarks can have Yukawa interactions $y$  with the elementary SM Higgs doublet $H$. 
The scenario that we consider is described by the renormalizable Lagrangian
\beq \Lag = \Lag_{\rm SM}^{m=0}  - \frac14 {\cal G}_{\mu\nu}^{A2}  +  \bar\Q_L^i i\slashed{D} \Q_L^i
 +  \bar\Q_R^j i\slashed{D} \Q_R^j  +( y_{ij}  H \Q_L^i \Q_R^j +\hbox{h.c.})
\eeq
where $\Lag_{\rm SM}^{m=0} $ describes the SM without the Higgs mass term,
and ${\cal G}^A_{\mu\nu}$ is the techni-color field strength. 
In models where Yukawa couplings $y$ are not allowed 
(for example techni-quarks in the 3 of $\SU(2)_L$) 
the number of free parameters is the same as in the Standard Model:
all new physics is univocally predicted.
This new physics manifests as:
\begin{itemize}

\item Strong dynamics generates a dynamical scale $\Lambda_{\rm TC}$ that can be identified with the mass of the lightest vector meson resonance,
the techni-$\rho$, and spontaneously breaks accidental chiral symmetries conserved by the techni-strong interactions producing 
light pseudo-Goldstone bosons (GB). Using large $N$ counting $m_\rho = g_\rho f$ where $f$ is the decay constant of the techni-pions and
 $g_\rho \approx 4\pi/\sqrt{N}$.

\item 
In absence of techni-quark masses, the techni-pions $\pi \sim \Q_L \Q_R$ acquire mass $m_\pi^2 \approx \alpha_{2} m_\rho^2/4\pi$
from the electro-weak gauge interactions that explicitly break the global techni-flavour accidental symmetries.
Yukawa couplings also contribute to their masses; in absence of Yukawa couplings the lightest techni-pions could be a stable 
$\SU(2)_L$ triplet providing a viable DM candidate.

\item The heaviest new particles are techni-baryons with mass $m_B \approx N m_\rho $. The lightest techni-baryon is stable and is 
a natural DM candidate; if it is  a thermal relic, the observed DM abundance is reproduced for $m_B \approx 100\TeV$~\cite{KamGr}.

\end{itemize}
The LHC phenomenology of techni-strong dynamics was discussed in~\cite{Sundrum}.
The main new point of our work is the possible connection with the weak scale and implications for dark matter.
Assuming that power divergences vanish~\cite{FN,agrav}, the techni-strong interactions give a finite {\em negative} contribution 
to the Higgs squared mass term, such that the weak scale is dynamically generated.
The Higgs physical mass arises as
\beq M_h^2 \approx + \alpha_2^2 f^2  + y^2 \frac{m_\rho^2 f^2}{m_\pi^2}\label{eq:stima}\eeq
so that
{\em the techni-color scale is predicted to be $f \approx M_h/\alpha_2 \approx \hbox{\rm few $\TeV$}$},
or smaller in models where $y$ is present and dominant in eq.\eq{stima}.
Unlike ordinary techni-color as a solution to the usual hierarchy problem, 
where the natural scale for new physics is the weak scale itself,
in this scenario the natural mass scales are
\beq m_\pi \sim 2\TeV,\qquad m_\rho \sim 20\TeV,\qquad m_B \sim 50\TeV.\eeq
New physics effects in accelerator searches and precision experiments are well below the present sensitivity.
In particular no new effects are generated in flavor physics.
Techni-pions~\cite{hill} and techni-baryons~\cite{Bbaryon}, 
stable due to accidental symmetries of the renormalizable Lagrangian, 
can provide a thermal Dark Matter candidate.

\medskip

This work is organised as follows.
In section~\ref{m} we consider the Higgs mass generated by the SM electro-weak gauge couplings,
by the SM strong coupling, and by the Yukawa couplings of the Higgs with the techni-quarks, allowed in some models.
Dark Matter is discussed in section~\ref{DM}. We conclude in section~\ref{concl}. In the appendix we present the 
technical details of the computation of the potential induced by Yukawa interactions.

\section{Higgs Mass}\label{m}

We write the tree-level potential of the SM Higgs doublet $H$ as
\beq V = m^2 |H|^2 + \lambda |H|^4 . \eeq
If  $m^2 \equiv - M_h^2/2$ is negative, the Higgs doublet $H$ develops the vacuum expectation value
$v = M_h/\sqrt{2\lambda}\approx 246.2\GeV$:
expanding the potential $V$ around its minimum as
$ H=(0,(v+h)/\sqrt{2})$ shows that  $M_h\approx 125\GeV$ is the tree-level mass of the physical Higgs boson $h$.

Under our assumptions, the only mass scale of the theory is set by the dynamical scale of the techni-color sector.
Through loop corrections it induces other scales and in particular the Higgs mass parameter.
Electro-weak interactions of the techni-quarks induce  a 2-loop contribution, computed in section~\ref{g2},
and  color charges give a 3 loop contribution to the Higgs mass, computed in section~\ref{g3}.
If the Higgs couples to the techni-quarks through Yukawa interactions (for example if techni-quarks
contain doublets and singlets under the electro-weak interactions) a contribution to the Higgs mass 
is also generated at 1-loop, computed in section~\ref{Y}.

\subsection{Electro-weak interactions}\label{g2}
\begin{figure}
$$\includegraphics[width=0.93\textwidth]{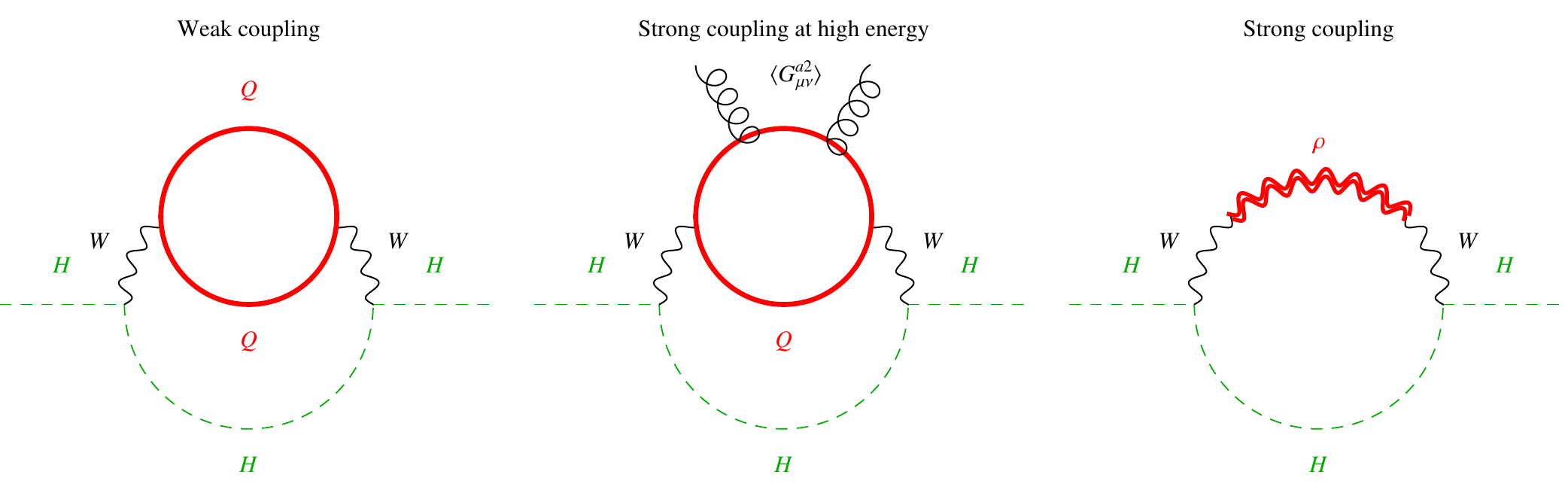}$$
\caption{\em 
\label{fig:gauge} The two loop contribution to the Higgs mass coming from the electro-weak gauge interactions of: 
a) a techni-quark, to be dressed with non-perturbative techni-interactions, approximated as:
b) the techni-gluon condensate;
c) the techni-$\rho$.  The extra seagull diagram is not explicitly plotted. }
\end{figure}

Electro-weak gauge interactions give a minimal, quasi-model-independent, contribution to the Higgs mass, described by the
non-perturbative techni-color multi-loop dressing of the two-loop Feynman diagram in fig.\fig{gauge}a
(plus the associated seagull diagram): the Higgs interacts with the electro-weak vectors, that interact with the techni-quarks.

\medskip

To leading order in the SM interaction, and to all orders in 
the techni-strong interactions, the techni-strong dynamics corrects the SM electro-weak gauge bosons propagator as
\bea   
D^{YY}_{\mu\nu}(q)&=& - i \frac{\eta_{\mu\nu} }{q^2}(1+g_Y^2\Pi_{YY}(q^2)) + i \xi_Y \frac{q_\mu q_\nu}{q^2}\\
D^{ab}_{\mu\nu}(q)&=& - i \frac{\eta_{\mu\nu} }{q^2}(1+g_2^2\Pi_{WW}(q^2)) \delta^{ab}+ i \xi_W \frac{q_\mu q_\nu}{q^2}\delta^{ab}
\eea
where $\xi_V$ are gauge-fixing parameters. 
Techni-strong dynamics is encoded in the $\Pi_{VV}(q^2)$ functions.
From the point of view of the techni-strong dynamics, they are the
renormalised two-point functions of the currents
$J_\mu^a = \sum_i  \bar \Q^i \gamma_\mu T^a_{\Q^i} \Q^i$
(where $\Q^i = (\Q^i_L,\bar\Q^i_R)$ is a Dirac spinor and $T^a$ are the SM gauge generators)
corresponding to the unbroken part of the accidental  global techni-flavour symmetry,
partially gauged by electro-weak interactions:
\begin{equation}
i\int d^4x\, e^{iq\cdot x}\langle 0| \hbox{T} J^V_\mu(x) J^{V'}_\nu(0)|0\rangle  \equiv  \delta^{VV'}(q^2 g_{\mu\nu}-q_\mu q_\nu) \Pi_{VV}(q^2).
\end{equation}
The correction to the Higgs mass is
\beq \label{eq:dmm1}
\Delta m^2 = - \frac{3}{4i}\int\frac{d^4q}{(2\pi)^4} \frac{3 g_2^4 \Pi_{WW}(q^2) + g_Y^4 \Pi_{YY}(q^2)}{q^2}  ,
\eeq
and, performing the Wick rotation to the Eucliedan $Q^2 = - q^2 >0$,
\beq \label{eq:dmm}
\Delta m^2  = 
\frac{3}{4(4\pi)^2} \int dQ^2  \bigg[3g_2^4 \Pi_{WW}(-Q^2)+ g_Y^4 \Pi_{YY}(-Q^2)\bigg].
\eeq
In general the integral above is UV-divergent, quadratically and logarithmically.
In the case at hand, the unphysical power divergences are ignored because of
our assumption of ñfinite naturalnessî, and logarithmic divergences (that describe the RGE running of $m^2$) are absent,
because of our assumption that the only mass scale, $\Lambda_{\rm TC}$, is generated dynamically.
Thereby the  generated squared Higgs mass term is finite and scheme independent.

\smallskip 
 
We next show that the electro-weak interactions induce a calculable negative Higgs mass so that the electro-weak symmetry is spontaneously broken.
We proceed in 3 steps: dispersion relations in section~\ref{disp} show in general that $\Delta m^2 <0$,
Operator Product Expansion in section~\ref{OPE} shows that $\Delta m^2$ is ultra-violet finite,
vector meson dominance and/or large $N$ in section~\ref{rho} allow to give the estimate $\Delta m^2 \approx - \alpha_2^2 f^2$.

\subsubsection{Dispersion relation}\label{disp}
Under our assumptions, quadratically divergent terms are zero and we are interested in the dependence 
on the physical scales of the theory. To extract this we consider the variation of the Higgs mass with respect
to the dynamical scale of the theory $\Lambda_{\rm TC}$,
\beq \label{eq:ddmm}
\frac{\partial \Delta m^2}{\partial \Lambda^2_{\rm TC}}   = 
\frac{3g_2^4}{4(4\pi)^2} \int dQ^2 \bigg[ 3g_2^4 \frac {\partial \Pi_{WW}}{\partial \Lambda^2_{\rm TC}}
+ g_Y^4  \frac {\partial \Pi_{YY}}{\partial \Lambda^2_{\rm TC}}\bigg].
\eeq
The sign of the gauge correction  $\Delta m^2$ can be determined using the dispersion relation~\cite{SVZ}
\beq
\frac {\partial \Pi_{VV}(q^2)}{\partial q^2} =  \frac{1}{\pi}\int_0^\infty ds \frac{{\rm Im}\, \Pi_{VV}(s)}{(s-q^2 -   i\epsilon)^2}.
\eeq
where we use the conventions of  \cite{Peskin}.
The optical theorem relates the cross-sections $\sigma(s)$ to ${\rm Im}\, \Pi_{VV}(s)$, allowing to show in general that ${\rm Im}\, \Pi_{VV}\le 0$.\footnote{As a check,
replacing  techni-color with a perturbative one-loop correction of fermions with explicit mass $m_\Q$, 
one would obtain
\beq\label{eq:dPiweak}
 \frac{\partial \Pi_{VV}(-Q^2)}{\partial m_\Q^2} = - \frac {g^2}{2\pi^2} \frac {Q^2}{m_\Q^2} \int_0^1 \frac {x^2(1-x)^2}{m_\Q^2+ x(1-x)Q^2}.
\eeq
Inserting this into eq. (\ref{eq:ddmm}) the integrand is negative definite but the integral is logarithmically divergent. 
This corresponds to a contribution proportional to $g^2 m_\Q^2$ in the
RG equation for the Higgs mass $m^2$.
No such UV-divergent 
RGE effect is present in a techni-color theory that
generates dynamically a mass scale $\Lambda_{\rm TC}$ from a dimension-less coupling $g_{\rm TC}$,
given that, in any mass-independent scheme such as Minimal Subtraction,
only $g_{\rm TC}$ can appear in the RGE.}
For dimensional reasons, the dimension-less $\Pi_{VV}$ can only depend on $Q^2/\Lambda^2_{\rm TC}$. Thereby
\beq
\frac{\partial \Pi_{VV}}{\partial \Lambda^2_{\rm TC}}=- \frac {Q^2}{\Lambda^2_{\rm TC}} \frac{\partial \Pi_{VV}}{\partial Q^2}
= \frac {Q^2}{\Lambda^2_{\rm TC}} \frac 1 {\pi} \int\frac {{\rm Im} \Pi_{VV}(s)}{(s+Q^2)^2}ds < 0
\eeq
where in the last step we used the dispersion relation.
A similar relation holds for the hypercharge contribution.
The integrand in (\ref{eq:ddmm}) is {negative} definite corresponding to a {negative} $\Delta m^2$
given the boundary condition $\Delta m^2=0$ for $\Lambda_{\rm TC}=0$.

\subsubsection{The ultra-violet tail}
\label{OPE}

In a theory with a dynamical scale $\Lambda_{\rm TC}$, arguments based on Operator Product Expansion 
allow to show that  $\partial \Delta m^2/\partial \Lambda^2_{\rm TC}$ is ultra-violet convergent as expected and to compute the
high-energy tail of $\Pi_{VV}(q^2)$.   $\Pi_{VV}$ can be expanded as
\beq \Pi_{VV}(q^2)  \stackrel{q^2 \gg \Lambda_{\rm TC}^2}\simeq
c_1 (q^2) + c_2(q^2) \langle0| m_\Q \Q_L \Q_R |0\rangle+
c_3(q^2)  \langle 0| \frac{\alpha_{\rm TC}}{4\pi}\G_{\mu\nu}^{A2}|0 \rangle+ \cdots  .
\label{eq:OPE}\eeq
The first term (unity operator) does not contribute to (\ref{eq:ddmm}). Indeed, at leading order it describes the diagram in fig.\fig{gauge}a with
techni-quarks but neglecting their techni-color interactions, such that 
\beq c_1 =   \frac{C}{12\pi^2}\ln(-q^2) + \cdots \eeq 
where $C>0$ is a model-dependent group theory factor given by $C =\Tr T^a T^a$ in terms of the $\SU(2)_L$ techni-quark generators
(with a similar expression factor for the U(1)$_Y$ generators).
% such that $C = N_F N n (n^2-1)/4$ for $N_F$ fundamentals
%of $\SU(N)$ in the $n$-dimensional representation of $\SU(2)_L$ (times 2 if it is complex?).
This high energy tail does not contain any mass scale, so that the associated quadratically divergent no-scale integral in eq.\eq{dmm} vanishes, under our assumptions.
The second term also vanishes, because  it is proportional to the techni-quark masses $m_\Q$
that vanish under our assumption that the theory does not contain any mass scale.

\medskip

The third term in eq.\eq{OPE} is  represented by the Feynman diagram in fig.\fig{gauge}b, which gives $c_3 = -C'/ q^4$~\cite{SVZ},
where $C'>0$ is another order one model-dependent group theory factor.
The techni-gluons form a positive condensate (the condensate is positive-defined in the Eucliedian path-integral~\cite{SVZ},
in agreement with QCD lattice computations)
\beq \langle 0 | \frac{\alpha_{\rm TC}}{4\pi} \G_{\mu\nu}^{A2}|0\rangle = \kappa\, \Lambda_{\rm TC}^4 . \eeq
where $\kappa>0$ is an order-one coefficient.
This allows to show that the UV contribution to the squared Higgs mass term is negative as expected:
\beq \Delta m^2|_{\rm UV} \simeq -\frac{3C'g_2^4}{4(4\pi)^2} \kappa \Lambda_{\rm TC}^4 \int_{Q^2_{\rm min}}^\infty \frac{dQ^2}{Q^4} \approx-
\alpha_2^2   \frac{\kappa \Lambda_{\rm TC}^4}{Q_{\rm min}^2}.
\eeq
The $1/Q_{\rm min}^2$ dependence on the artificial infra-red cut-off $Q_{\rm min}\sim \Lambda_{\rm TC}$
shows that the dominant effects comes from virtual momenta $Q^2$ around the techni-meson masses.

\medskip

\subsubsection{The infra-red and resonance region}\label{rho}
The dominant contribution to the Higgs mass comes from the $Q^2$ region 
densely populated by the techni-meson resonances.
A variety of methods have been proposed to approximatively describe such region:
vector meson dominance, Weinberg sum rules, large $N$, holographic models...
As long as the techni-quarks are charged under the electro-weak group, 
they form, among the various mesons, spin-1 resonances that mix with the SM electro-weak vectors $V_\mu$.
This is described by the effective Lagrangian
\beq \Lag_{\rm eff}=- \frac{1}{4g_0^2} V_{\mu\nu}^a V^{a\,\mu\nu} -\frac 1 {4g_\rho^2}\rho_{\mu\nu}^a \rho^{a\,\mu\nu} + \frac {f^2}2 (V_\mu^a - \rho_\mu^a)^2\eeq
such that the massless  eigenstate has gauge coupling
$1/g^2_2 = 1/g_0^2 + 1/g_\rho^2$ and the orthogonal heavy state has mass
$m_\rho^2 = f^2(g_0^2  + g_\rho^2)$. 
Integrating out the $\rho$ at tree-level one finds:
\begin{equation}\label{eq:PiVV}
\Pi_{VV}(q^2)=\frac {m_\rho^2}{g_\rho^2(q^2-m_\rho^2+i \epsilon)}.
\end{equation}
Plugging eq.\eq{PiVV} into eq.~(\ref{eq:dmm}) we obtain a logarithmically divergent infra-red correction
to the squared Higgs mass term:
\begin{equation}
\Delta m^2 \approx - \frac{9g_2^4}{4(4\pi)^2} \int dQ^2 \frac {m_\rho^2}{g_\rho^2(Q^2+m_\rho^2)}
 \sim - \frac{g_2^4m_\rho^2}{(4\pi)^2 g_\rho^2}\log \frac{\Lambda^2}{m_\rho^2} \sim - \alpha_2^2\,f^2\  .\eeq
The integrand is negative definite and its size agrees with the naive expectation
based on the Feynman diagram plotted in fig.\fig{gauge}c,
including the $1/g_\rho^2$ suppression of vector mixing. The logarithmic UV divergence here 
arises because this is only an approximate description, where an explicit mass term $m_\rho$
substitutes the dynamical mechanism of mass generation.  An infinite number of  
states would be needed to properly describe the non-perturbative dynamics.

In theories with large $N$ this can be made more rigorous:
$\Pi_{VV}$ can be represented exactly as an infinite sum of poles 
corresponding to the physical quasi-stable techni-mesons of the theory:
\begin{equation}
\Pi_{VV}(q^2) =  \frac {N}{16\pi^2} m_\rho^2 \sum_i \frac {c_i^2}{q^2- m_i^2 + i \epsilon} .
\end{equation}
where $c_i$ are adimensional coefficients. 
The infinite number of resonances allows to reproduce the
logarithmic divergence, that does not contribute to the Higgs mass zero under our assumption of finite naturalness.

\smallskip

These considerations offer an intuitive argument to understand the sign of $\Delta m^2$.
The net effect of non-perturbative dynamics is creating a mass gap
that stops the techni-quark contribution to the RGE running of $g_2, g_Y$
below $\Lambda_{\rm TC}$, effectively making $g_2,g_Y$ smaller with respect to the perturbative case.
As a consequence the unphysical power divergence present in the SM,
$\Delta m^2 \sim +g_{2,Y}^2 \Lambda^2$,
gets replaced by a finite physical  effect $\Delta m^2 \sim - g_{2,Y}^4 \Lambda_{\rm TC}^2$.

\begin{figure}
$$\includegraphics[width=0.93\textwidth]{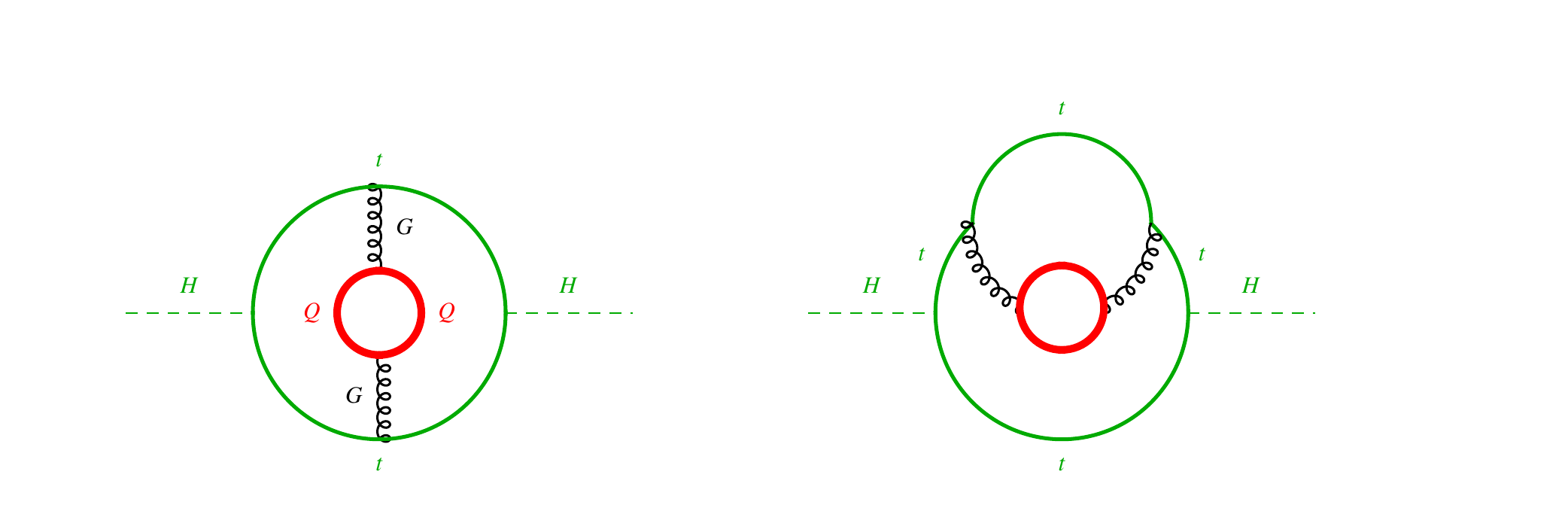}$$
\caption{\em 
\label{fig:strong} The three loop contribution to the Higgs mass coming from  
techni-quarks $\Q$ that only have color interactions. Similar diagrams can be drawn for graviton contributions.}
\end{figure}

\subsection{Color interactions}\label{g3}
We next consider techni-color models where the techni-quarks have SM color interactions.
For example, techni-quarks could be a color octet of $\SU(3)_c$, charged also under the techni-color gauge group.
Then techni-quarks cannot have any Yukawa coupling to the SM Higgs: both the Yukawa contribution of section~\ref{y1} 
and the electro-weak contribution of section~\ref{g2} are absent.

In these models, the Higgs mass is dominantly generated at three loops: 
the Higgs interacts with the top quark, that interacts with the gluons, that interact with the techni-quarks,
as plotted in fig.\fig{strong}.
The computation can be performed along the lines of section~\ref{g2} by defining $\Pi_{GG}(q^2)$, 
the techni-color correction to the gluon propagator. 
Summing the two diagrams of fig.\fig{strong}, the result is ultraviolet-convergent:
\begin{equation}
\Delta m^2= - \frac {64 y_t^2 g_3^4}{(4\pi)^4}\int dQ^2 \Pi_{GG}(-Q^2) \sim \frac{y_t^2 g_3^4}{4\pi^4} f^2.
\end{equation}
The computation of the sign is analogous to what described in the previous section (with $\Pi_{WW}$ replaced by $\Pi_{GG}$):
in the present case we find a positive $\Delta m^2$, such that this contribution does not induce
electro-weak symmetry breaking.
The sign of the effect also corresponds to the intuitive reasoning presented at the end of the previous section:  
the sign is opposite to the known negative sign of the naive quadratic divergence associated with $y_t$, because
$g_3$ and thereby $y_t$ are reduced by techni-strong dynamics.

\bigskip

We mention a final possibility. The techni-quarks  could be completely neutral under the whole  SM gauge group.
In this situation only gravity mediates a contribution to the Higgs mass, proportional to the  two-point function 
of the energy momentum tensor. Furthermore, a super-Planckian techni-color condensate would dynamically generate
the Planck mass itself, within a dimensionless extension of Einstein gravity such as agravity~\cite{agrav}.
The problem is that techni-color dynamics, dominated by a single non-perturbative coupling,
has no free parameters and would also generate a large negative cosmological constant,
which is at odd with observations.

\medskip

\subsection{Yukawa interactions}\label{Y}\label{y1}

Finally, we consider the case where the gauge quantum numbers of the techni-quarks
allow for Yukawa couplings to the elementary Higgs. This choice implies the existence of a
techni-pion $\pi_2$  with the same quantum numbers of the Higgs doublet $H$, that can then mix with $H$.

The left panel of fig.\fig{Yuk} shows the one-loop corrections to the squared Higgs mass generated by a 
weakly coupled techni-quark with Yukawa interactions to the Higgs. 
At strong coupling the physical degrees of freedom become bound state techni-hadrons that can be described using effective Lagrangian techniques.
The  techni-quark loop can be matched to an effective chiral Lagrangian, so that such diagrams collapses to a tree level diagram 
(right-handed panel of fig.\fig{Yuk}) dominated by the lightest techni-mesons, the techni-pions $\pi \approx  {\cal Q}_L{\cal Q}_R$.
For simplicity we here consider Yukawa couplings that preserve the $\Q_L\leftrightarrow\Q_R$ parity
of the techni-strong interactions; a more general discussion can be found in the appendix. 
Similarly to quark masses in QCD, the Yukawa interactions produce the following term in the chiral Lagrangian,
\begin{equation}
y\, m_\rho f^2\,  {\rm Tr}[H U] +\hbox{h.c.}
\end{equation}
where  $U= \exp(i \pi^{\hat{a}} T^{\hat{a}}/f)$ is the Goldstone boson matrix. As we discuss in detail in the appendix, upon minimisation of the potential 
this term induces a mass mixing $\approx y m_\rho f H \pi^*$ between the techni-pion and the elementary Higgs.
This term also explicitly breaks accidental symmetries respected by gauge interactions.

\begin{figure}
$$\includegraphics[width=0.93\textwidth]{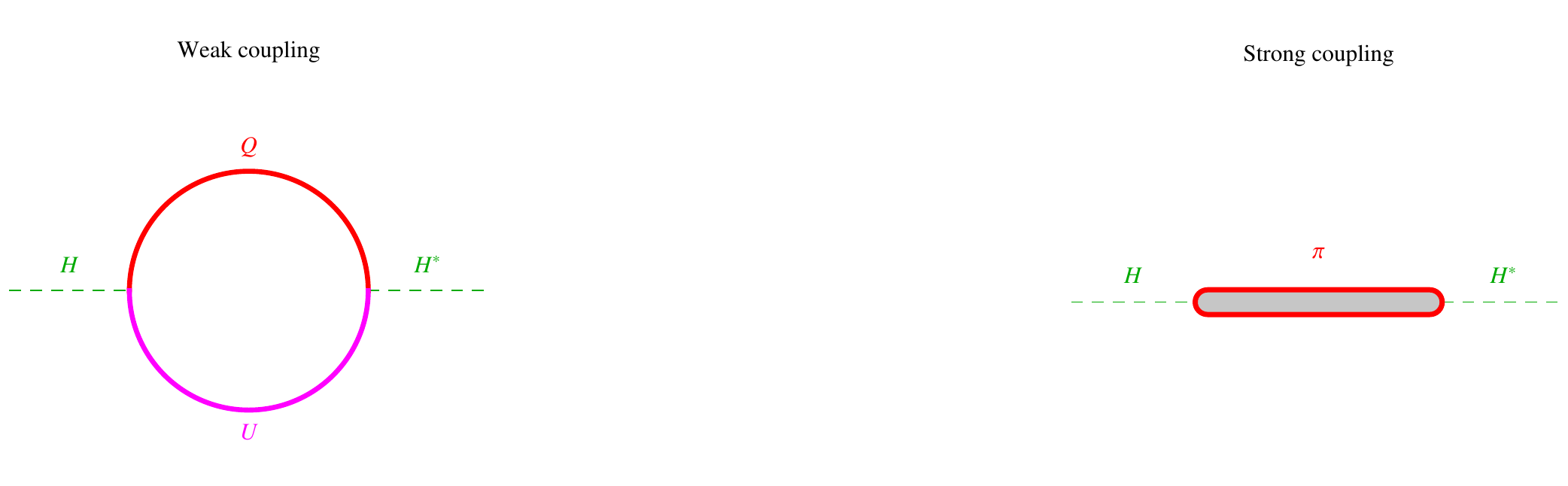}$$
\caption{\em 
\label{fig:Yuk} Correction to the Higgs mass coming from the Yukawa coupling with: a)
a weakly coupled massive fermion; b) a massless strongly interacting fermion.}
\end{figure}

What emerges is a two-Higgs doublet system where the extra Higgs doublet $\pi_2$ is a heavy composite doublet with negligible vev.  
In order to compute the mass eigenstates, we need to compute the mass matrix.
Including effects at tree and  one-loop level in the SM couplings $g_2$ and $y$, the mass matrix has the structure 
\beq\bordermatrix{ & \pi_2^* & H^* \cr 
\pi_2 &  ({\cal O}(g_2^2) \pm {\cal O}(y^2))/(4\pi)^2 & {\cal O}(y) \sqrt{N}/(4\pi) \cr 
H &  {\cal O}(y) \sqrt{N}/(4\pi) &-{\cal O}( y^2)  N/(4\pi)^2}  \label{pihmatrix}  m_\rho^2
\label{eq:mpiH}
 \eeq
where we used the fact that the one-loop contribution of weak gauge interactions to $m_\pi^2 \approx g_2^2 m_\rho^2/(4\pi)^2$
is  positive (as known from the SM analogous computation of the $\pi^+$/$\pi^0$ mass difference~\cite{witten}),
and added the one-loop Yukawa contribution (absent in the SM\footnote{The literature on composite Higgs models 
explored linear couplings of SM quarks to composite fermionic states, finding that  they can give a negative contribution
to the Higgs mass term. Simple UV completions require extra scalars as in the supersymmetric realisation of~\cite{Ser2}.
Here instead we compute the  techni-pion potential induced by a bi-linear $H \Q_L\Q_R$ Yukawa coupling,
involving techni-quarks $\Q$ and a scalar $H$ without techni-strong interactions.}).
The $HH^*$ entry describes the contribution of composite scalar resonances that can also mix with the 
Higgs giving a negative sub-leading contribution to its mass squared, see appendix for more details. 

We see that the phenomenologically acceptable regime is the one where the Yukawa coupling is small,
$y \ll  g $,  such that:
1) the loop contribution coming from the Yukawa coupling can be ignored;
2) the heaviest eigenstate is the techni-pion with squared mass $m_\pi^2>0$;
3) the determinant of the mass matrix is dominated by the off-diagonal terms and is negative: the lightest eigenstate is the elementary Higgs, 
that acquires a {\em negative} squared mass term dominated by the mass mixing term in eq.\eq{mpiH} and
given by a see-saw-like formula:
\beq \Delta m^2 \sim  - \frac {y^2}{(4\pi)^2} \frac{m_\rho^4 N}{m_\pi^2}\sim - y^2 \frac{m_\rho^2 f^2}{m_\pi^2}\, .\eeq

\section{Dark Matter}\label{DM}

The models described in this paper contain two Dark Matter (DM) candidates: techni-baryons and techni-pions. 
Their stability is guaranteed by accidental symmetries of the renormalizable Lagrangian, techni-baryon 
number and (possibly) $G$-parity \cite{hill}.

In fact the presence of stable states is a generic prediction of the framework that implies restrictions on the representations of the
techni-quarks under the SM gauge group, such that the stable states are viable DM candidates.
In table~\ref{dmsummary} we summarise the simplest allowed charge assignments 
under the electro-weak  group and  the resulting DM candidates. Introducing techni-quark masses allows
several other possibilities \cite{compositeDM}.  

\smallskip

The new matter modifies the running of SM gauge couplings. Adding 
$n_2$ weak doublets and $n_3$ weak triplets in the $N\oplus\bar{N}$ of $\SU(N)_{\rm TC}$
the beta-function of $\SU(2)_L$ becomes
\beq b_2 = - \frac{19}{6} + \frac{2N}{3} (n_2 +4 n_3) \eeq
such that the $\SU(2)_L$ gauge coupling does not develop a Landau pole below the Planck scale
($b_2 \circa{<}5$) and possibly remains asymptotically free ($b_2<0$) for small enough $n_2,n_3,N$. 
Higher $SU(2)_L$ lead to Landau poles instead. The trans-Planckian Landau pole for hypercharge can be naturally avoided in models where 
hypercharge is embedded in $\SU(2)_R$ below a few TeV~\cite{Landau}; a technicolor sector 
could be used to dynamically break the extended gauge group.

\begin{table}[t]
$$\begin{array}{lc|cc|cc|l}
\multicolumn{2}{c|}{ \hbox{number of}}
&  \multicolumn{2}{c|}{N=3} &  \multicolumn{2}{c|}{N=4} \\  
\multicolumn{1}{c}{ \hbox{techni-flavors}}&\hbox{Yukawa}&
 \hbox{TCb} & \hbox{TC}\pi  & \hbox{TCb}  & \hbox{TC}\pi\\
\hline
% \rowcolor[rgb]{0.95,0.95,0.95}   N_F =   1 & 1 &1\\  \hline
\rowcolor[cmyk]{0,0,0.2,0.1}
\multicolumn{1}{c}{ N_F=2}  & & {2} & 3  & {1}&3 & \hbox{under TC-flavor $\SU(2)$} \\
 \rowcolor[cmyk]{0,0,0.2,0}
%  1_{Y}+1_{Y'} &2\times \pi^{Y Y'}+\textcolor{red}{\pi^0}&2\times \pi^{Y Y'} +\textcolor{red}{\pi^0}\\  
\hbox{model 1: } \Q=2_{Y=0}&0& \hbox{charged} & \hbox{3} &{1} & 3  & \hbox{DM, under $\SU(2)_L$}\\   \hline
 \rowcolor[cmyk]{0,0.2,0,0.1}
\multicolumn{1}{c}{N_F=3}& &  {8} & 8 & { \bar 6} & 8 & \hbox{under TC-flavor $\SU(3)$}  \\ 
 \rowcolor[cmyk]{0,0.2,0,0}
%1_{Y_1} + 1_{Y_2} +1_{Y_3}&6\times \pi^{Y_i Y_j} +\textcolor{red}{\pi^0}&6\times \pi^{Y_i Y_j}+\textcolor{red}{\pi^0} \\
\hbox{model 1: }\Q=1_Y + 2_{Y'}&1& {1} & \hbox{no} & {1} & \hbox{no} & \hbox{DM, under $\SU(2)_L$}\\
 \rowcolor[cmyk]{0,0.2,0,0}
\hbox{model 2: } \Q=3_{Y=0}&0 & {3}&3  &{1} &3 & \hbox{DM, under $\SU(2)_L$}\\  \hline
\rowcolor[cmyk]{0.1,0,0.1,0.1}
\multicolumn{1}{c}{N_F=4} &&{ \overline{20}}&15& {20'}&15& \hbox{under TC-flavor $\SU(4)$}\\
 \rowcolor[cmyk]{0.1,0,0.1,0}
%2_0+2_0& 1^- &-&6\times  1_0 +2\times (1^-_{\pi}+\pi_{IJ}^0)\\
%1_0 +3_0 &1^-,2^+&1_0+1^-_{\pi}+2\times \pi_{IJ}^0&6\times  1_0 +1^-_{\pi}+2\times \pi_{IJ}^0\\
\hbox{model 1: }\Q=4_{Y=0}&0& \hbox{charged}&3 & { 1}&3 & \hbox{DM, under $\SU(2)_L$}\\
\rowcolor[cmyk]{0.2,0,0,0.1}
\multicolumn{1}{c}{N_F=5} &&{ \overline{40}}&24& {\overline{50}}&24& \hbox{under TC-flavor $\SU(5)$}\\
 \rowcolor[cmyk]{0.2,0.0,0,0.0}
%2_0+2_0& 1^- &-&6\times  1_0 +2\times (1^-_{\pi}+\pi_{IJ}^0)\\
%1_0 +3_0 &1^-,2^+&1_0+1^-_{\pi}+2\times \pi_{IJ}^0&6\times  1_0 +1^-_{\pi}+2\times \pi_{IJ}^0\\
\hbox{model 1: }\Q=2_Y + 3_{Y'}&1& 1&\hbox{no} & \hbox{charged}&\hbox{no} & \hbox{DM, under $\SU(2)_L$}\\
 \rowcolor[cmyk]{0.2,0.0,0,0.0}
\hbox{model 2: }\Q=5_{Y=0}&0& 3&3 & 1&3 & \hbox{DM, under $\SU(2)_L$}
\end{array}
$$
\caption{\em Dimension-less techni-color models that give viable techni-baryon (TCb) and/or
techni-pion (TC$\pi$) Dark Matter candidates with $Q=Y=0$. We consider models with $\SU(N)$ gauge group for $N=\{3,4\}$
and $N_F=\{2,3,4,5\}$ flavours of techni-quarks in its fundamental plus anti-fundamental.
The darker rows give the  techni-flavour content of the lightest TCb and TC$\pi$ considering only masses induced by techni-color interactions.
The lighter rows  consider models with viable assignments of  electro-weak interactions and show, after including the 
mass splitting due to unbroken electro-weak interactions, the $\SU(2)_L$ content of the lighter DM candidates. 
\label{tab:DM1}}
\label{dmsummary}
\end{table}%J=1

\subsection{Techni-pions}\label{G}

If techni-quarks fill $N_F$ fundamentals and anti-fundamentals of the $\SU(N)_{\rm TC}$ gauge group with $N\ge 3$, 
the  spontaneous symmetry breaking $\SU(N_F)_L\otimes \SU(N_F)_R/\SU(N_F)$ of the accidental
global techni-flavor symmetry produces $N_F^2-1$ Goldstone bosons in the adjoint of the unbroken $\SU(N_F)$. These scalars acquire mass
from effects that explicitly break the global symmetries. Within finite naturalness the only contribution to 
their masses is due to SM gauge interactions, and possibly to the techni-quark Yukawa couplings.

If Yukawa couplings are forbidden by the fermions quantum numbers, then the model is extremely predictive:
it only has one free parameter --- the techni-color scale --- which is fixed by the Higgs mass under the
hypothesis of finite naturalness.  All the rest is univocally predicted: techni-pion masses, Dark Matter and
its thermal relic abundance.

The SM gauge interactions give positive squared masses to the gauge-charged techni-pions, 
while  SM singlets remain exact massless Goldstone bosons.  If the $N_F$ techni-quark
flavors are composed of $k$ irreducible (real or pseudo-real) representations of $G_{\rm SM}$, 
then the techni-pions decompose under $G_{\rm SM}$ as
\begin{equation}
{\rm Adj}_{\SU(N_F)}= \left[\sum_{i=1}^k r_i \right]\otimes \left[\sum_{i=1}^k  \bar{r}_i \right] \ominus 1
\end{equation}
so that $k-1$ techni-pions are neutral gauge singlets (the extra scalar singlet analog of the $\eta'$ in QCD 
acquires mass from anomalies with techni-interactions and will not play a role in what follows).

One combination of singlets corresponds to a global symmetry anomalous under $\SU(2)_L$,
so that the corresponding Goldstone boson acquires an axion-like couplings to SM vectors:
an almost massless axion with a decay constant  $f\sim \TeV$ would be grossly excluded by star cooling and other bounds.
In absence of techni-quark Yukawa interactions, these bounds significantly reduce the space of models
favouring the simplest models with $k=1$. The techni-quarks should belong to a single irreducible representation $j=(N_F-1)/2$ of $\SU(2)_L$  
and, in order to obtain a neutral lightest techni-baryon, the techni-quark hypercharge should vanish.
Then the $N_F^2-1$ techni-pions lie in the following irreducible representations $J$ of $\SU(2)_L$:
\beq {\rm Adj}_{\SU(N_F)}=\sum_{J=1}^{N_F-1} J . \eeq 
Models of this kind were studied in \cite{hill}, where it was pointed out that a discrete symmetry, 
``$G$-parity" exists in these theories (for zero hypercharge)
due to the fact that $\SU(2)_L$ representations are real or pseudo-real. $G$ parity  acts on techni-quarks as
$\Q\to \exp (i \pi T^2)\Q^c$, replacing any $\SU(2)_L$ representation with 
its conjugate representation, which is equivalent to the original representation.
SM fields are neutral.
On techni-pions $G$ parity becomes the  $(-1)^J$  Z$_2$ symmetry, 
so that techni-pions with even (odd) isospin (J) are even (odd). 
Summarizing:
\begin{itemize}
\item Techni-pion singlets under $\SU(2)_L$ are $G$-even, do not acquire masses from SM gauge interactions and can have
anomalous couplings to $\SU(2)_L$ vectors: they are excluded in our framework unless Yukawa couplings make them massive.
They are absent if techni-quarks fill a single irreducible representation of $\SU(2)_L$.

\item Techni-pions in the 3 of $\SU(2)_L$ are $G$-odd and could be the lightest stable DM candidates.
The simplest models are listed in table~\ref{tab:DM1}.

\item Techni-pions in the 5 of $\SU(2)_L$ are $G$-even and are heavier,
$m_{\pi_5}\approx \sqrt{3} m_{\pi_3}$: they undergo anomalous decays into electro-weak vectors,
$\pi_5 \to WW$.

\item Techni-pions in higher representations of  $\SU(2)_L$, if present, decay into lighter techni-pions respecting $G$-parity
by emitting two $\SU(2)_L$ vectors, e.g.\ $\pi_7 \to \pi_3 WW$.

\end{itemize}

\bigskip

The situation is different in models where Yukawa couplings $y$ of techni-quarks to the elementary
Higgs are present. The Yukawa couplings break explicitly $G$-parity and accidental global symmetries
so that the SM singlet techni-pions $\eta$ receive non-zero masses given by eq.~(\ref{meta}),
$M_\eta \sim |y| v m_\rho/m_{\pi_2}$ and star cooling bounds are easily avoided.
Furthermore, techni-pions can now decay through the Higgs, so that only techni-baryons remain as dark matter candidates.

\bigskip

Models with techni-color gauge group  $\SU(2)\sim  {\rm Sp}(2)$ are
special: its fundamental representation is pseudo-real, $2\sim 2^*$,
so that the techni-flavour symmetry is enhanced becoming $\SU(2 N_F)/{\rm Sp}(2 N_F)$. The extra techni-pions 
are $\Q\Q$ scalars and there are no stable  techni-baryons. 
Dangerous light techni-pions neutral under $\SU(2)_L$ are again absent if 
techni-quarks lie in a single representation of $\SU(2)_L$ with dimension  $2 N_F$.
Within our assumptions however these models do not provide DM candidates because techni-pions 
are $G$-even.

\subsection{Techni-baryons}\label{TCB}
Techni-baryons are techni-color singlet states constructed with $N$ techni-quarks. The stability of
the lightest techni-baryon follows from the accidental 
techni-baryon number global symmetry. 

Using the non-relativistic quark model, group theory allows to compute the  electro-weak quantum numbers of the techni-baryons:
their wave-function must be anti-symmetric in the techni-quarks. The wave function is assumed to be antisymmetric in techni-color, and so 
must be symmetric in spin and flavour for the lightest techni-baryons that have no orbital angular momentum.
Different techni-baryons are split by spin-spin interactions that prefer, as lightest techni-baryon, the one with smallest spin. 
As a consequence, the lightest techni-baryons have  spin 0 (1/2)  for even (odd) $N\ge2$.

In general the $\SU(N_F)$ techni-flavour representation of the lightest techni-baryon corresponds to a Young diagram with  2 rows
having $N/2$ boxes each (for $N$ even)
and to a Young diagram with 2 rows having $(N+1)/2$ and  $(N-1)/2$ boxes respectively (for $N$ odd).
 In particular, they are
\begin{equation}
\raisebox{-1.5ex}{\yng(2,1)} \qquad\hbox{for $N=3$}\qquad\hbox{and}\qquad  
\raisebox{-1.5ex}{\yng(2,2)} \qquad \hbox{for $N=4$}.
\end{equation}
This is the end of the story, as long as techni-color interactions are involved.

\medskip

Next, the components of a techni-baryon multiplet are split by SM gauge interactions,
and possibly by techni-quark Yukawa interactions. The lightest components are those with smallest $G_{\rm SM}$ charge. 

Furthermore, electro-weak symmetry breaking induces extra splitting 
within the components of any electro-weak multiplet,
with the result that the component with smallest electric charge  is the lightest stable state~\cite{MDM}. 
Since DM direct detection constraints demand that DM does not couple at tree level to the $Z$, 
the DM hypercharge should be zero, which is possible for integer isospin.

\subsection{Direct detection of Dark Matter}\label{dd}
The previous discussion is  summarised in table~\ref{tab:DM1}, which tells 
that the simplest TC models lead to the following viable stable DM candidates: 
\begin{itemize}
\item  Techni-baryons, fermions for odd $N$ and scalars for even $N$.
Their annihilation cross section is estimated to be $\sigma v \sim g_{\rm TC}^4/4\pi M^2$,
around the unitarity bound~\cite{KamGr}.
By performing a naive rescaling of the QCD non-relativistic $p\bar p$ cross section,
$\sigma_{p\bar p} v \sim 100/m_p^2$, we
estimate that the cosmological thermal relic abundance of a techni-baryon
equals the total DM abundance 
if its mass is loosely around $m_B\sim 200\TeV$.
A cosmological techni-baryon asymmetry can leave a higher abundance, allowing for a lighter $m_B$.

\item Scalar techni-pions,
that fill a $\SU(2)_L$ triplet with hypercharge $Y=0$.
Techni-pions have small residual techni-color interactions (as well as small quartic couplings)
and thereby behave as Minimal Dark Matter~\cite{MDM}.
Their cosmological thermal relic abundance equals the total DM abundance 
if their mass is around $2.5 \TeV$~\cite{MDM}.
Their spin-independent cross section for direct detection is
$\sigma_{\rm SI} \approx 0.12~10^{-46}\cm^2$~\cite{Hisano,FN},
as plotted in fig.\fig{DM}.
\end{itemize}
As already discussed,
both  mass scales suggested by the DM cosmological abundance arise naturally within the context of finite naturalness.

\begin{figure}[t]
$$\includegraphics[width=0.7\textwidth]{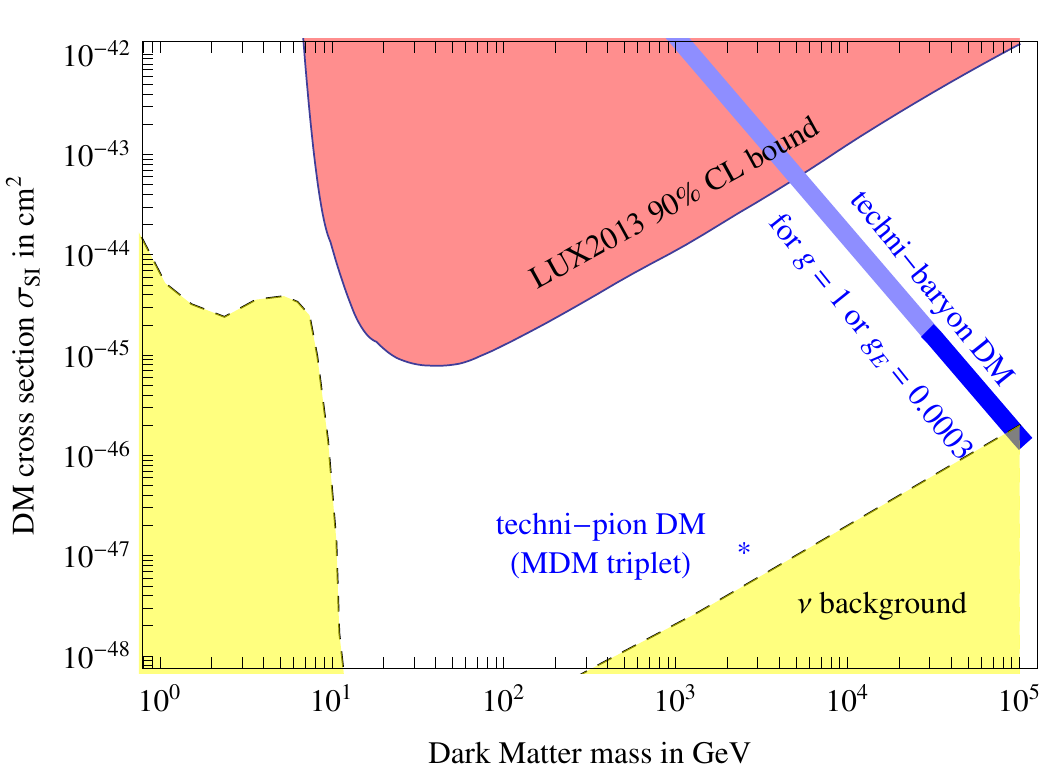}$$
\caption{\em 
\label{fig:DM} The signals in Dark Matter direct detection produced by a DM techni-baryon with magnetic or electric dipole moment
(line) or from a Minimal-Dark-Matter-like techni-pion with thermal abundance (star), compared to the present experimental LUX bound ~\cite{LUX} and to the background due to neutrinos. }
\end{figure}

\bigskip

Techni-baryons have distinctive features in direct detection experiments: 
if DM is a neutral composite particle made of charged techni-quarks, 
direct detection can be mediated by  the photon~\cite{dip}.
Any such DM particle can have a non trivial form factor,
dominated at low energy by the `charge radius' interaction.
For a scalar DM $S$ this is the only interaction and can be written as
\beq
\frac{e}{\Lambda_{\rm TC}^2} (S^* i \partial_\alpha S) \partial_\mu F^{\mu \alpha}.
\eeq
The resulting  cross section for direct detection is suppressed by four powers of the TC scale,
and is negligible for $\Lambda_{\rm TC}\sim\hbox{few TeV}$.

The situation is more promising if  DM is a fermionic techni-baryon $B$, which generically has magnetic (and possibly
electric) dipole moments, $\mu$ and $d$.
They are described by the effective operator
\beq    \bar B \sigma_{\mu \nu} \frac{\mu + i d  \gamma_5}{2} B ~   F_{\mu\nu} .\eeq
Electro-magnetic dipoles give sizeable
direct detection signals with a characteristic testable enhancement at low recoil-energy $E_R$,
given that the DM/matter scattering is mediated by the massless photon.
Furthermore, in the relevant non-relativistic limit, the cross-section induced by
the magnetic dipole $\mu$ is suppressed by two extra power of the relative
DM/matter velocity $v$ with respect to the cross section induced by the more speculative electric dipole $d$~\cite{dip}
\beq \frac{d\sigma}{dE_R} \approx \frac{ e^2 Z^2}{4\pi  E_R}\left(\mu^2 + \frac{d^2}{v^2}\right) .
\eeq
For simplicity, we here assumed a nucleus with $Z\gg1$, a recoil energy $E_R \ll m_{\cal N} v^2$
and approximated the nuclear charge form factor with unity.

\smallskip

The magnetic $g$-factor, defined by  $ \mu = g e/2m_B$, is expected to be of order one
for a strongly-coupled particle (while it is loop suppressed for an elementary particle).
We also define the electric $g$-factor as  $d = g_E  e/2m_B$.
In terms of such $g$-factors we find that the present direct detection bound is
\beq  g^2 +  1.2~10^7  g_E^2  < \bigg( \frac{m_B}{5.1\TeV}\bigg)^3
\label{LUXMD}
\eeq
dominated by LUX data~\cite{LUX,Cirelli}. 
This bound assumes that  techni-baryons constitute all galactic DM, and must be rescaled otherwise.
Fig.\fig{DM} shows the resulting prediction in the usual plane ($M_{\rm DM}, \sigma_{\rm SI}$) used
to describe spin-independent direct detection of Dark Matter.

An electric dipole moment needs CP-violation.
In our context, techni-quarks are strictly massless, such that the CP-violating  techni-strong $\theta$ term is not physical.
A small $g_E$ could be generated if techni-quark masses are included.

\subsection{A worked example}\label{example}
More quantitative predictions can be given in the QCD-like scenario with $N=N_F=3$~\cite{Bbaryon}. 
In this case the spectrum can be obtained by rescaling known QCD results,
\beq
\frac{m_B}{m_{\rho}} \approx 1.3 \qquad \qquad \frac{m_{\pi}}{m_{\rho}} \approx 0.1\, \sqrt{J(J+1)}
\eeq
where $m_\rho$ is the mass of the lightest techni-vector resonance and techni-pions $\pi$ lie in the $J$
representation of $\SU(2)_L$. The second estimate is obtained from the electro-magnetic
splitting of QCD pions, see the appendix.

The lightest techni-baryons are an octet of flavour $\SU(3)$ and table~\ref{tab:DM1} lists the two possible viable 
assignments for techni-quarks under $\SU(2)_L\otimes{\rm U}(1)_Y$:
\beq \Q = \left\{\begin{array}{l}
1_{\mp1/3} \oplus 2_{\pm1/6}
\cr
3_{0}
\end{array}\right. .\eeq
The hypercharges are determined by requiring that the lightest techni-baryon is neutral;
in the first case their overall normalisation is determined by
requiring that techni-quarks can have a Yukawa interaction with the Higgs. 
For this choice of quantum numbers the lightest techni-baryon is an electro-weak singlet with $Y=Q=0$,
avoiding direct detection constraints.

\smallskip

The lightest technibaryons decompose under $\SU(2)_L$ as
\beq
\mb{8}=\left\{\begin{array}{ll}
\mb{2}({\rm p,n}) \oplus\mb{3}(\Sigma^{\pm,0})\oplus\mb{2}(\Xi^0,\Xi^-)\oplus\mb{1}(\Lambda_0) & 
\hbox{for $\Q=1 \oplus 2$} \\
\mb{3} \oplus \mb{5}& \hbox{for $\Q=3$} \  
\end{array}\right. \ .
\eeq
In the $\Q=1\oplus 2$ model  we used the familiar names of the QCD octet. 
The lightest techni-baryon is $\Lambda_0$, that is 
analogous to the QCD state $\Lambda_0^{\rm QCD} \sim s(ud-du)$. 
Its magnetic dipole moment can be estimated from QCD data:
$\mu_{\Lambda_0^{\rm QCD}}=0.61 e/2m_p$~\cite{PDG}.
Inserting $g=-0.61$ in eq.~(\ref{LUXMD}) we obtain the bound $m_{\Lambda^0}>3.7$ TeV.
The previous QCD-based estimate of the DM annihilation cross section becomes exact, such that 
the cosmological DM density is reproduced for $m_{\Lambda^0}\approx 200\TeV$.
In this model there are no stable techni-pions.

\medskip

In the $\Q=3_Y$ model the lightest technibaryon is a triplet $\mb{3}_{3Y}$ of $\SU(2)_L$,
so that neutral DM is obtained for $Y=0$ and $Y=\pm 1/3$: the first possibility is allowed by direct detection constraints.
Due to the absence of  Yukawa  and hypercharge interactions, 
the neutral member of the techni-pion triplet is the DM candidate, stable thanks to the accidental 
G-parity discussed in section~\ref{G}.
Its mass must be smaller than $2.5\,{\rm TeV}$ in order to avoid a thermal relic density
bigger than the observed DM density.
This implies that in this model the thermal relic density of the technibaryon dark matter is subdominant.

\section{Conclusions}\label{concl}

In conclusion, we presented a new class of models where the Standard Model
is made dimension-less by dropping the  mass term of the elementary Higgs
and extended by adding  techni-quarks with techni-color interactions arranged in such a way that
they do {\em not} break the electro-weak gauge group nor generate a composite Higgs.
Within the context of finite naturalness ---
the assumption that a QFT with no mass parameters nor power divergences
might provide a revised concept of weak-scale naturalness and of the origin of mass scales ---
the simplest models of this type dynamically generate a mass term for the Higgs.

The elementary Higgs acquires a squared mass term $m^2$ entirely determined 
in terms of weak interactions of the techni-quarks and of the techni-color scale.
Using various approximation techniques that allow to control the techni-color dynamics, 
in section~\ref{g2} we found
that the sign of $m^2$ is negative, such that $\SU(2)_L\otimes{\rm U}(1)_Y$ gets broken,
and that the observed weak scale is obtained for a techni-color scale $m_\rho \approx 4\pi M_h /\alpha_2 \approx 10-20 \TeV$.
This is large enough that such models do not pose any phenomenological problem.
Techni-pions are lighter, as determined by their electro-weak interactions, and could give observable signals at LHC; in particular
techni-pions $\pi_5$ in the 5 of $\SU(2)_L$ undergo anomalous decays into pairs of electro-weak vectors,
$\pi_5\to WW$.
Such models can have the same number of free parameters as the Standard Model:
{\em all new physics is univocally predicted}, up to theoretical uncertainties
in the techni-strong dynamics, that could be reduced with respect to our estimates
by performing  dedicated lattice computations.\footnote{Techni-strong dynamics generates a negative vacuum energy of order $- \Lambda_{\rm TC}^4$.
It can be canceled, compatibly with the scenario of dynamical mass generation in the SM sector,
by adding another sector negligibly coupled to SM particles; this kind of sector is anyhow needed to account for the Planck mass.
This cancellation is the usual huge fine-tuning  associated with the cosmological constant problem, on which we have nothing to say.}

\medskip

Independently of the assumption of finite naturalness, the models studied in this paper contain two Dark Matter candidates:
the lightest techni-baryon $B$ with mass $m_B \sim 50\TeV$ (section~\ref{TCB}) and, in some models, the lightest techni-pion $\pi_3$, a triplet under $\SU(2)_L$ with mass
$m_{\pi_3} \sim 0.1 m_\rho\sim 1-2~ \TeV$ (section~\ref{G}).
Their thermal relic abundance is also univocally predicted, with the result that
the observed cosmological Dark Matter abundance is naturally reproduced in the techni-pion case,
while the techni-baryon seems more likely to be a sub-dominant Dark Matter component,
if a naive rescaling of the QCD $p\bar p$ cross-section holds, and ignoring possible techni-baryon asymmetries.
The direct detection cross section of such DM candidates
is predicted to be $2-3$ orders of magnitude below present bounds.
Magnetic moment interactions of techni-baryons would lead to recoil events with a distinctive energy spectrum (section~\ref{dd}).

\medskip

Table~\ref{tab:DM1} offers a panoramic of models that lead to DM candidates.
In some models the quantum numbers 
allow for Yukawa interactions between techni-quarks and the elementary Higgs.
Such Yukawas give extra negative contributions to the squared Higgs mass term
(section~\ref{Y}), so that the techni-color scale needed to reproduce the weak scale gets lighter;
in such models a singlet techni-pion is especially light.
Models where techni-quarks only have QCD interactions or gravitational interactions 
do not seem to lead to a promising phenomenology, as discussed in section~\ref{g3}.

\small

\subsubsection*{Acknowledgments}
We wish to thank Roberto Franceschini and Giovanni Villadoro for discussions and collaboration at the early stages
of this work. We thank Roberto Contino, Riccardo Rattazzi, Slava Rychkov, Riccardo Torre for
advice about strong dynamics and the authors of~\cite{Cirelli} (in particular Eugenio del Nobile) for help about using their code.
The work of OA and MR is supported by the MIUR-FIRB grant RBFR12H1MW.
%A contract signed with Rattazzi grants him the right of adding a sentence to this paper: ``...''.

\appendix

\section{Effective potential}
The  effective potential for the elementary Higgs and the techni-pions receives contributions at tree level in the Yukawa couplings
and at loop level in the gauge and Yukawa couplings. It  can be computed using the techniques reviewed in~\cite{contino}:
the relevant ingredients are the correlation functions of the composite operators of the theory. 
There are three main contributions: from SM gauge interactions at loop level (section~\ref{Vg1});
from the possible Yukawa couplings at tree level (section~\ref{Vy0})
and at loop level (section~\ref{Vy1}).
Summing these contributions, the full potential is studied in section~\ref{Vsum}.

\subsection{Gauge  contribution at one loop level}\label{Vg1}

At 1-loop the SM gauge interactions 
induce a techni-pion mass that can be computed in terms of correlators of
the vector ($J_\mu^a = \sum_\Q  \bar \Q \gamma_\mu T^a_\Q \Q$) and axial ($J_\mu^a = \sum_\Q  \bar \Q \gamma_\mu T^{\hat{a}}_\Q \gamma_5 \Q$) symmetry currents.  On general grounds these have the form,
\begin{eqnarray} 
 i\int d^4x\, e^{iq\cdot x}\langle 0| \hbox{T} J^V_\mu(x) J^{V'}_\nu(0)|0\rangle& \equiv  \delta^{VV'}(q^2 g_{\mu\nu}-q_\mu q_\nu) \Pi_{VV}(q^2),\nonumber \\
 i\int d^4x\, e^{iq\cdot x}\langle 0| \hbox{T} J^{A}_\mu(x) J^{A'}_\nu(0)|0\rangle & \equiv  \delta^{AA'}(q^2 g_{\mu\nu}-q_\mu q_\nu) \Pi_{AA}(q^2).
\label{2ptcurrent}
\end{eqnarray}
The one-loop techni-pion potential reads~\cite{contino}:
\begin{equation}
V_{g1}\approx \frac 3 {2(4\pi)^2} \sum_i g_i^2 {\rm Tr}[U T^i U^\dagger T^i] \int_0^\infty Q^2 dQ^2  \left[\Pi_{AA}(-Q^2)- \Pi_{VV}(-Q^2)\right] 
\end{equation}
where $U= e^{i \pi^{\hat{a}} T^{\hat{a}}/f}$ is the Goldstone boson matrix,
$g_i$ are the SM couplings and $T_i$ their generators. 
Gauge-charged techni-pions acquire positive squared masses, that, for the $\SU(2)_L$ interactions, are estimated as
\begin{equation}
m_\pi^2\approx \frac {3g_2^2}{(4\pi)^2} J (J+1) m_\rho^2
\end{equation}
where $J$ is the weak isospin of the techni-pion representation.

\medskip

\subsection{Yukawa contribution at tree level}\label{Vy0}
We now consider the potential generated by the Yukawa interactions.
For concreteness we here focus on the case where techni-quarks $\Q=2\oplus 1$ fill 
one doublet and one singlet of $\SU(2)_L$ with hypercharges as in section \ref{example}. The 8 techni-pions  decompose under $\SU(2)_L \otimes {\rm U}(1)_Y$ as
\begin{equation}
8 = 2_{ \pm 1/ 2} +3_0 + 1_0 .
\end{equation} 
In general there are two Yukawa couplings:
\beq  y H \Q^1_L\Q^2_R + y'  H^\dagger \Q^1_R \Q^2_L + \hbox{h.c.}
= H \bar\Q_2 \bigg(\frac{y+y'^*}{2} +\gamma_5 \frac{-y + y'^*}{2}\bigg) \Q_1+\hbox{h.c.}
\eeq
where on the right hand side we used Dirac spinors $\Q_i = (\Q^i_L,\bar \Q^i_R)$. 
The phases of $y$ and $y'$ are not physical 
and can be chosen for convenience, for example real and positive. The terms above generate the tree level
effective potential
\begin{equation}
V_{y0}=a_0 {\rm Tr}[M U] +\hbox{h.c.}
\end{equation}
where $a_0\approx - m_\rho f^2$ and
\begin{equation}
\label{spurion}
M= 
\bordermatrix{
& \Q_R^1  & \Q_R^2\cr
\Q_L^1 &0  & y H \cr
%0 & 0 & y h_0 \\
\Q_L^2 & y' H^\dagger  & 0 
}\, .
\end{equation}
The explicit result for the potential of the doublet ($\pi_2$) and singlet ($\eta$) techni-pions is
\begin{equation}
\label{pionpot}
V_{y0}= -8\sqrt{2}  a_0 \, {\rm Im} \, e^{-\frac{i\,\eta}{4\sqrt{3}f}}\, \frac{\sin \Delta/f}{\Delta} \,[y H^\dagger \pi_2 + y' \pi_2^\dagger H],\qquad\Delta=\frac 1 4\sqrt{3 \eta^2+8 \pi_2^\dagger \pi_2}.
%V_{y0}= i\, 4\, a_0 \, e^{-\frac{i\,\eta}{4\sqrt{3}f}}\, \frac{\sin \Delta/f}{\Delta} \,[y H^\dagger \pi_2 + y' \pi_2^\dagger H]+\hbox{h.c.}\,,\qquad\Delta=\frac 1 4\sqrt{3 \eta^2+8 \pi_2^\dagger \pi_2}.
\end{equation}

\subsection{Yukawa contribution at one loop level}\label{Ay1}\label{Vy1}

To compute the one-loop Yukawa correction to the effective potential we proceed 
similarly to the gauge interactions. We formally
introduce sources $S_{\bar{i}j}$ for the techni-quark bilinears $\Q_L^i \Q_R^{\bar{ j}}(x)$
(such that, in the real theory of interest, it contains $yH$ in some of its components)
and write the effective Lagrangian that describes the Higgs/techni-pion system after having integrated 
out the heavier techni-strong dynamics. For simplicity we consider vectorial couplings as in these case
fewer invariants exist. In a constant techni-pion configuration and  to quadratic order 
in the sources $S$, the effective action has the following structure determined by the symmetries,
\begin{equation}
\Lag_{\rm eff}^{QQ}= a_0 \delta^4(q) ({\rm Tr}[S U]+\hbox{h.c.}) + \Pi_0^{QQ}(q^2)  {\rm Tr}[S S^\dagger]+\Pi_1^{QQ}(q^2)|{\rm Tr}[S U]|^2.
\label{effyuk}
\end{equation}
The first term linear in $S$ describes the $\Q_L\Q_R$ condensate.
The form factors can be obtained integrating over the strong dynamics including techni-pion fluctuations.
By construction they  encode the two point functions of the techni-quark bilinears,
\begin{equation}
\langle 0| \bar{\Q}^i \Q^{\bar{j}} (q)\, \bar{\Q}^{\bar{k}} \bar{\Q}^l(-q)|0\rangle=i\, G^{QQ}_{\rm Adj}(q^2) \left(\delta^{i\bar{k}}\delta^{l\bar{j}}-\frac 1 3 \delta^{i\bar{j}}\delta^{l\bar{k}}\right)+i\,G^{QQ}_S(q^2) \delta^{i\bar{j}}\delta^{l\bar{k}}
\label{QQ2pt}
\end{equation}
where $G^{QQ}_S$ and $G^{QQ}_{\rm Adj}$ correspond to the singlet and adjoint channels
(namely, the octet for $N_F=3$). 
Matching eq.s~(\ref{effyuk}) and (\ref{QQ2pt}) (for example choosing $U= {\rm 1}$) one finds
\begin{equation}
\Pi_0^{QQ}=G_{\rm Adj}^{QQ}\,,\qquad \Pi_1^{QQ}=G_S^{QQ}-\frac 1 3 G_{\rm Adj}^{QQ} .
\end{equation}
At large $N$ one has
\begin{equation}
G_{\rm Adj}^{QQ}(q^2)=\frac {N}{16\pi^2} m_\rho^4 \sum_n \frac {c_{{\rm Adj}_n}^2}{q^2-m_{{\rm Adj}_n}^2+i\epsilon}\,,   
\qquad
G_S^{QQ}(q^2)= \frac {N}{16\pi^2} m_\rho^4 \sum_n \frac {c_{S_n}^2}{q^2-m_{S_n}^2+i\epsilon}\,,
\label{QQlargeN}
\end{equation}
where the coefficients $c$ are of order 1 and the sum is over the scalar resonances in the theory.
The sum does not include techni-pions because we only consider vectorial Yukawa couplings that do not generate 1 techni-pion states.

To obtain the effective action for the scalars we just need to set to zero the non dynamical components of $S$ 
and add kinetic terms for the components of $S$  associated to the Higgs. This produces
\begin{equation}
\Lag_{\rm eff}^{\rm H} = a_0 \delta^4(q)  ({\rm Tr}[M U]+ \hbox{h.c.})+ (q^2+ y^2 \Pi_0^{QQ}(q^2)) H^\dagger H+ \Pi_1^{QQ}(q^2) |{\rm Tr}[M U]|^2.
\end{equation}
The first term describes the tree level contribution discussed above. The second term encodes the tree level effect of mixing
with heavy scalar resonances that gives the $HH^*$ entry  of the mass matrix in eq.~(\ref{eq:mpiH}),
\begin{equation}
m_H^2=y^2 \Pi_0^{QQ}(0)\sim - \frac {y^2\,N}{(4\pi)^2}m_\rho^2.
\end{equation}
Performing the path integral with respect to $H$ we obtain the one-loop Yukawa contribution to the techni-pion potential, 
\begin{eqnarray}
V_{y1}& =&\frac 1 2 \int \frac {d^4 Q}{(2\pi)^4}\ln \left[Q^2-y^2 \Pi_0^{QQ}(-Q^2) - y^2 \Pi_1^{QQ}(-Q^2)\sum_{a} {\rm Tr}[T^a U]{\rm Tr}[T^a U^\dagger]\right]\nonumber \\
&\approx &v_0^4- \frac {y^2} {2(4\pi)^2}\sum_{a} {\rm Tr}[T^a U]{\rm Tr}[T^a U^\dagger] \int_0^\infty dQ^2  \Pi_1^{QQ}(-Q^2)
\label{ypotential}
\end{eqnarray}
where $v_0$ is the contribution to the the vacuum energy and $T^a$ are $\SU(3)$ matrices derived from (\ref{spurion}). 

One can prove that, similarly to the gauge contribution, the loop integral in~(\ref{ypotential}) is  finite:
since $\Pi_1^{QQ}$ is sensitive to the chiral symmetry breaking, the Operator Product Expansion demands that
\begin{equation}
\Pi_1^{QQ}(q^2)\stackrel{q^2 \gg \Lambda_{\rm TC}^2}\simeq
\frac {\langle 0| (\bar{\Q}_L \Gamma_1 \Q_R)(\bar{\Q}_R \Gamma_2\Q_L) |0 \rangle}{q^4}
\end{equation}
where $\Gamma_{1,2}$ are appropriate matrices in techni-color and flavour space, see \cite{SVZ}. 

Contrary to the gauge contribution we are not aware of any theorem that guarantees the sign of this contribution. 
As an estimate the contribution above gives
\begin{equation}
\delta m_\pi^2 \sim \frac {y^2m_\rho^2}{(4\pi)^2} .
\end{equation}
Summing up all the contributions we obtain a mass matrix with the structure of eq.~(\ref{pihmatrix}).

\subsection{Minimization of the potential}\label{Vsum}
The vacuum is determined through the minimization of the potential
\beq V_{\rm eff}(\pi, \eta, H)=  V_{g1} +V_{y0}+ V_{y1} +m^2 |H|^2 + \lambda |H|^4\eeq
where $m^2<0$ is induced by gauge loops  (section~\ref{g2}). 
The gauge-charged techni-pions $\pi$ acquire a large mass from gauge loops and can be integrated out,
leaving an effective potential for the lighter scalars: the elementary Higgs doublet $H$ and the gauge-neutral techni-pion $\eta$.
In the parameter range of interest for us, $g\gg y$,  one has $V_{y1}\approx0$ and $V_{g1}\approx \frac12 m_{\pi}^2 \pi^2(1-\eta^2/16 f^2)$, where, for simplicity, we expanded at second order in $\eta/f$ sufficient to compute the mass of the singlet.
We can freely redefine the phases of the Yukawa couplings $y$ and $y'$ so that $y y'$ is real and negative. 
With this choice  $\eta=0$ indeed is a local minimum of the effective potential
\begin{equation}
V_{\rm eff}(\eta\ll f, H) \approx |H|^2 \bigg[ m^2 -32 \, \frac{m_\rho^2 f^2}{m_\pi^2}\left((|y|+|y'|)^2-|y y'|\frac {\eta^2}{12 f^2}\right)\bigg]+\lambda|H|^4.
%V_{\rm eff}(\eta\ll f, H) \approx |H|^2 \bigg[ m^2 -128 |y|^2 \, \frac{m_\rho^2 f^2}{m_\pi^2}\left(1-\frac {\eta^2}{48 f^2}\right)\bigg]+\lambda|H|^4.
\end{equation}
Around the minimum $\eta$ acquires a positive squared mass 
\begin{equation}
\label{meta}
M_\eta\sim |y| \frac {m_\rho}{m_\pi} v
\end{equation}
without mixing with the Higgs, that receives
a negative contribution to its $m^2$ parameter. 
%To see this note that from eq. (\ref{pionpot}) it follows that $\eta=0$ is an extremum
%of the potential when the couplings have equal phases. 


\footnotesize

\begin{thebibliography}{nn}\bibitem{Sundrum} C.~Kilic, T.~Okui and R.~Sundrum,
  %``Vectorlike Confinement at the LHC,''
  JHEP {1002}, 018 (2010)
  [\hhref{0906.0577}].

\bibitem{FN}
 M.~Farina, D.~Pappadopulo and A.~Strumia,
  %``A modified naturalness principle and its experimental tests,''
  JHEP {1308}, 022 (2013)
  [\hhref{1303.7244}].
   A.~de Gouvea, D.~Hernandez and T.~M.~P.~Tait,
  %``Criteria for Natural Hierarchies,''
  Phys.\ Rev.\ D {89} (2014) 115005
  [\hhref{1402.2658}].

\bibitem{Bardeen}
F. Englert, C. Truffin and R. Gastmans, Nucl. Phys. B 117 (1976) 407.
W Bardeen, \href{http://lss.fnal.gov/archive/1995/conf/Conf-95-391-T.pdf}{FERMILAB-CONF-95-391-T}.
C.~T.~Hill,
  %``Conjecture on the physical implications of the scale anomaly,''
  \hhref{hep-th/0510177}.
 
\bibitem{agrav}
  A.~Salvio and A.~Strumia,
  %``Agravity,''
  JHEP {1406} (2014) 080
  [\hhref{1403.4226}].
  
\bibitem{prev}   
R. Hempfling,
 %``The Next-to-minimal Coleman-Weinberg model,''
  Phys.\ Lett.\ B {379} (1996) 153
  [\hhref{hep-ph/9604278}].
   J.~P.~Fatelo, J.~M.~Gerard, T.~Hambye and J.~Weyers,
  %``Symmetry breaking induced by top loops,''
  Phys.\ Rev.\ Lett.\  {74} (1995) 492.
  T.~Hambye,
  %``Symmetry breaking induced by top quark loops from a model without scalar mass,''
  Phys.\ Lett.\ B {371} (1996) 87
  [\hhref{hep-ph/9510266}].
  W.~-F.~Chang, J.~N.~Ng and J.~M.~S.~Wu,
  %``Shadow Higgs from a scale-invariant hidden U(1)(s) model,''
  Phys.\ Rev.\ D {75} (2007) 115016
  [\hhref{hep-ph/0701254}].
  R.~Foot, A.~Kobakhidze and R.~R.~Volkas,
  %``electro-weak Higgs as a pseudo-Goldstone boson of broken scale invariance,''
  Phys.\ Lett.\ B {655} (2007) 156
  [\hhref{0704.1165}].
R.~Foot, A.~Kobakhidze, K.~L.~McDonald, R.~R.~Volkas,
  %``A Solution to the hierarchy problem from an almost decoupled hidden sector within a classically scale invariant theory,''
  Phys.\ Rev.\ D {77} (2008) 035006
  [\hhref{0709.2750}].
    S.~Iso, N.~Okada, Y.~Orikasa,
  %``The minimal B-L model naturally realized at TeV scale,''
  Phys.\ Rev.\ D {80} (2009) 115007
  [\hhref{0909.0128}].
  S.~Iso and Y.~Orikasa,
  %``TeV Scale B-L model with a flat Higgs potential at the Planck scale - in view of the hierarchy problem -,''
  PTEP {2013} (2013) 023B08
  [\hhref{1210.2848}].
C.~Englert, J.~Jaeckel, V.~V.~Khoze and M.~Spannowsky,
 %``Emergence of the electro-weak Scale through the Higgs Portal,''
  \hhref{1301.4224}.
   E.J.~Chun, S. Jung, H.M.~Lee,
  %``Vacuum Stability, Perturbativity, EWPD and Higgs-to-diphoton in Type II Seesaw,''
  \hhref{1304.5815}.
  T.~Hambye and A.~Strumia,
  %``Dynamical generation of the weak and Dark Matter scale,''
  Phys.\ Rev.\ D {88} (2013) 055022
  [\hhref{1306.2329}].
  C.~D.~Carone and R.~Ramos,
  %``Classical scale-invariance, the electro-weak scale and vector dark matter,''
  Phys.\ Rev.\ D {88} (2013) 055020
  [\hhref{1307.8428}].
   R.~Foot, A.~Kobakhidze, K.~L.~McDonald and R.~R.~Volkas,
  %``Poincare Protection for a Natural electro-weak Scale,''
  Phys.\ Rev.\ D {89} (2014) 115018
  [\hhref{1310.0223}].
    A.~Farzinnia, H.~J.~He and J.~Ren,
  %``Natural Electroweak Symmetry Breaking from Scale Invariant Higgs Mechanism,''
  Phys.\ Lett.\ B {727} (2013) 141
  [arXiv:1308.0295].
  C.~T.~Hill,
  %``Is the Higgs Boson Associated with Coleman-Weinberg Dynamical Symmetry Breaking?,''
  Phys.\ Rev.\ D {89} (2014) 073003
  [\hhref{1401.4185}.
  J.~Guo and Z.~Kang,
  %``Higgs Naturalness and Dark Matter Stability by Scale Invariance,''
  \hhref{1401.5609}.
  S.~Benic and B.~Radovcic,
  %``electro-weak breaking and Dark Matter from the common scale,''
  Phys.\ Lett.\ B {732} (2014) 91
  [\hhref{1401.8183}].
   H.~Davoudiasl, I.M.~Lewis, 
  %``Right-Handed Neutrinos as the Origin of the Electroweak Scale,''
  Phys.\ Rev.\ D {90}, 033003 (2014)   [\hhref{1404.6260}].
K.~Allison, C.~T.~Hill and G.~G.~Ross,   \hhref{1404.6268}.
  %``Ultra-weak sector, Higgs boson mass, and the dilaton,''
   G.~M.~Pelaggi,
  %``Predictions of a model of weak scale from dynamical breaking of scale invariance,''
  \hhref{1406.4104}. 
  W.~Altmannshofer, W.~A.~Bardeen, M.~Bauer, M.~Carena and J.~D.~Lykken,
  %``Light Dark Matter, Naturalness, and the Radiative Origin of the electro-weak Scale,''
  \hhref{1408.3429}.
  
  


\bibitem{Raidal}
T.~Hur and P.~Ko,
  %``Scale invariant extension of the standard model with strongly interacting hidden sector,''
  Phys.\ Rev.\ Lett.\  {106}, 141802 (2011)
  [\hhref{1103.2571}].  
  M.~Heikinheimo, A.~Racioppi, M.~Raidal, C.~Spethmann and K.~Tuominen,
  %``Physical Naturalness and Dynamical Breaking of Classical Scale Invariance,''
  Mod.\ Phys.\ Lett.\ A {29} (2014) 1450077
  [\hhref{1304.7006}]. 

\bibitem{Lindner}  
T.~Hambye and M.~H.~G.~Tytgat,
  %``Confined hidden vector dark matter,''
  Phys.\ Lett.\ B {683} (2010) 39
  [\hhref{0907.1007}].  
  M.~Holthausen, J.~Kubo, K.~S.~Lim and M.~Lindner,
  %``electro-weak and Conformal Symmetry Breaking by a Strongly Coupled Hidden Sector,''
  JHEP {1312} (2013) 076
  [\hhref{1310.4423}].
  See also~\cite{LindnerQCD}.
  
\bibitem{compositeDM}
Work in progress. 
 
\bibitem{KamGr}
  K.~Griest and M.~Kamionkowski,
  %``Unitarity Limits on the Mass and Radius of Dark Matter Particles,''
  Phys.\ Rev.\ Lett.\  {64} (1990) 615.
   B.~von Harling and K.~Petraki,
  %``Bound-state formation for thermal relic dark matter and unitarity,''
  \hhref{1407.7874}.
  
\bibitem{Landau}
  G.~F.~Giudice, G.~Isidori, A.~Salvio and A.~Strumia,
  %``Softened Gravity and the Extension of the Standard Model up to Infinite Energy,''
\hhref{1412.2769}.
 
\bibitem{hill} 
  Y.~Bai and R.~J.~Hill,
  %``Weakly Interacting Stable Pions,''
  Phys.\ Rev.\ D {82}, 111701 (2010)
  [\hhref{1005.0008}].
  
    \bibitem{Bbaryon}
  R.~Pasechnik, V.~Beylin, V.~Kuksa and G.~Vereshkov,
  %``Vector-like technineutron Dark Matter: is a QCD-type techni-color ruled out by XENON100?,''
  Eur.\ Phys.\ J.\ C {74} (2014) 2728
  [\hhref{1308.6625}].
  
  
\bibitem{witten}
E.~Witten,
  %``Some Inequalities Among Hadron Masses,''
  Phys.\ Rev.\ Lett.\  {51}, 2351 (1983).
  
\bibitem{Ser2}
 D.~Marzocca, A.~Parolini and M.~Serone,
  %``Supersymmetry with a pNGB Higgs and Partial Compositeness,''
  JHEP {1403} (2014) 099
  [\hhref{1312.5664}]. 
  
\bibitem{contino}
 R.~Contino,
  %``The Higgs as a Composite Nambu-Goldstone Boson,''
  \hhref{1005.4269}.

\bibitem{SVZ}
M.A. Shifman, A.I. Vainshtein, V.I. Zakharov, Nucl. Phys. B147 (1979) 385.

\bibitem{Peskin}
 See e.g.\ M. Peskin, D.V. Schroeder, ``An Introduction to Quantum Field Theory'', pag.\ 618.
Our $\Pi_{VV}$ is defined with the same sign as $\Pi$ in this book.

\bibitem{MDM}
M.~Cirelli, N.~Fornengo and A.~Strumia,
  %``Minimal dark matter,''
  Nucl.\ Phys.\ B {753} (2006) 178
  [\hhref{hep-ph/0512090}].
   M.~Cirelli, A.~Strumia and M.~Tamburini,
  %``Cosmology and Astrophysics of Minimal Dark Matter,''
  Nucl.\ Phys.\ B {787} (2007) 152
  [\hhref{0706.4071}].

\bibitem{Hisano}
  \art[1007.2601]{J.~Hisano, K.~Ishiwata and N.~Nagata}{Phys.\ Rev. D}{82}{115007}{2010}.
  %``Gluon contribution to the dark matter direct detection,''
  R.~J.~Hill and M.~P.~Solon,
  %``Universal behavior in the scattering of heavy, weakly interacting dark matter on nuclear targets,''
  Phys.\ Lett.\ B {707} (2012) 539
  [\hhref{1111.0016}].
  
\bibitem{LUX}
LUX Collaboration,
  %``First Dark Matter Search Results from the Large Underground Xenon (LUX) Experiment,''
  \hhref{1405.5906}.
  
\bibitem{dip}
See e.g.\
 V.~Barger, W.~Y.~Keung and D.~Marfatia,
  %``Electromagnetic properties of dark matter: Dipole moments and charge form factor,''
  Phys.\ Lett.\ B {696} (2011) 74
  [\hhref{1007.4345}].

\bibitem{Cirelli}
   E.~Del Nobile, M.~Cirelli and P.~Panci,
  %``Tools for model-independent bounds in direct dark matter searches,''
      	  \hhref{1307.5955}.
	 


\bibitem{PDG}
Particle Data Group Collaboration,
  %``Review of Particle Physics (RPP),''
  Chin.\ Phys.\ C {38} (2014) 090001.
 
\bibitem{LindnerQCD}
   J.~Kubo, K.~S.~Lim and M.~Lindner,
  %``electro-weak Symmetry Breaking by QCD,''
  \hhref{1403.4262}.
  In the past other authors considered a similar idea, but without an elementary Higgs: 
  W. J. Marciano, Phys. Rev. D21 (1980) 2425. 
  D. Lust, E. Papantonopoulos, K. Streng, and G. Zoupanos, Nucl. Phys. B268 (1986) 49.  
  
\end{thebibliography}
\end{document}